\documentclass[twocolumn,twoside,slac_two]{revtex4}
\usepackage{graphicx}
\usepackage{fancyhdr}
\begin{document}

\title{Challenges in Accelerator Beam Instrumentation}

%

\author{M.~Wendt}
\affiliation{Fermi National Accelerator Laboratory, Batavia, IL 60510, USA}

\begin{abstract}
The challenges in beam instrumentation and diagnostics for present and
future particle accelerator projects are presented.
A few examples for advanced hadron and lepton beam diagnostics are given.
\end{abstract}

\maketitle

\thispagestyle{fancy}


\section{Motivation or \newline ``Why Beam Instrumentation?''}
Any modern particle accelerator requires three core elements
for beam acceleration: guide fields, accelerating fields and vacuum.
The emphasis of every new accelerator project lies on these mission 
critical areas and the related technical components:
\begin{itemize}
 \item Guide fields
 \begin{itemize}
  \item Magnets -- dipoles, quadrupoles, sextupoles, other multipoles,...
  \item Correction / steering magnets
  \item Power supplies
  \item Cooling water, technical interlocks, ...sometimes cryogenics
 \end{itemize}
 \item{Accelerating fields(\$\$\$)}
  \begin{itemize}
   \item Cavities, waveguides, couplers
   \item Klystrons, modulators, PFNs, HV-supplies
   \item Interlocks, control systems, and again sometimes cryogenics!
  \end{itemize}
  \item{Vacuum}
  \begin{itemize}
   \item Pipes, pumps, flanges, etc., ...and a very clean environment!
  \end{itemize}
 \end{itemize}
While most of the beam characteristics are defined by these three elements,
it would be difficult, perhaps impossible to verify the beam properties and
further improve its quality without a sufficient set of beam diagnostics 
and instruments.
These specialized beam instruments are the ``ears'' and ``eyes'' of the accelerator,
and they provide only to way ``watch'' the beam, and allow to characterize its 
properties and quality.

During the design and construction phase of a typical particle accelerator project 
the focus is on the technical elements providing guide and accelerating fields,
particular if superconducting technologies are involved.
Once the machine hardware is in completed and the accelerator has to be commissioned
with beam, the focus shifts to beam properties and quality, and the related beam instrumentation.
During this beam commissioning phase, but also after major upgrades or modifications,
a substantial set of well understood beam diagnostics is very important.
The instruments will help to spot errors (e.g. cabling, polarity, timing signals) and 
component failures (RF, magnets, power supplies, etc.).
But the beam instruments itself may not be fully operational from day one, and often need
test beam for internal tests and verification, thus the beam commissioning of the
machine and the related beam diagnostics have to move forward hand in hand.

Once the beam instruments are fully operational and well understood in their
performance and limitations, they can by used in a systematical way.
The beam parameter(s) of interest are extracted from the measured signals -- typically with help of 
digital signal processing technologies -- acquired, collected, stored and sorted through
the control system.
Now the accelerator beam diagnostics is used to analyze and characterize the beam, and
help to improve the properties and quality to the needs of the users.

The beam instrumentation can be seen as a detection element of a complex feedback system,
with, and for some damping or orbit FB systems even without human integration.
The measured beam parameters are compared with the simulations and give a better understanding
of the accelerator hardware and beam dynamics, thus the core input to modify, improve and
upgrade accelerator components and systems.
Therefore, improvements of the beam quality or sophisticated /exotic beam properties require
improvements and advances in the technology of the beam diagnostics.

\begin{figure}[h]
\centering
\includegraphics[width=80mm]{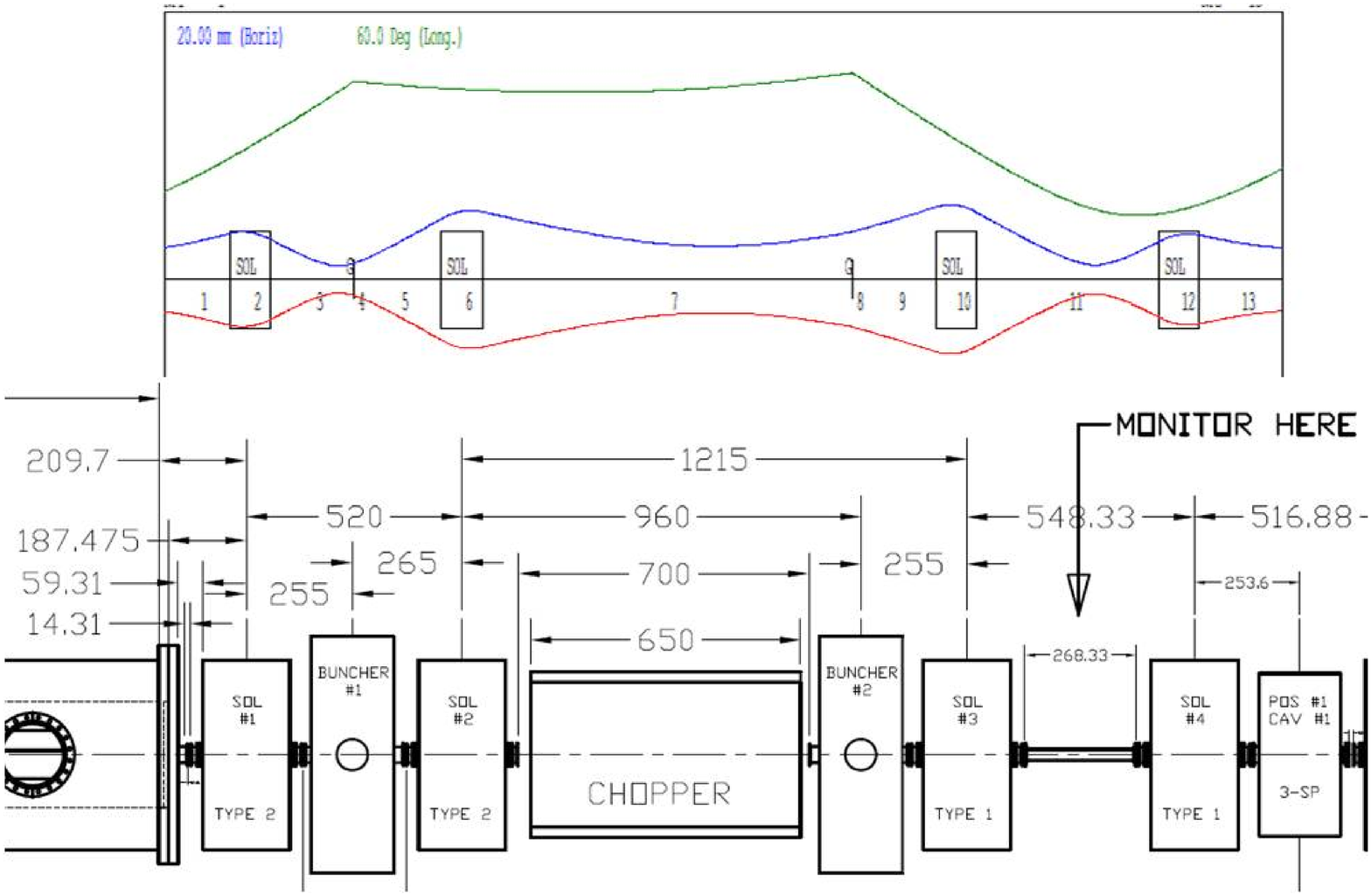}
\caption{``Don't forget beam diagnostics!''.} \label{forget-diag_figure}
\end{figure}
Fig.~\ref{forget-diag_figure} shows the discrepancy between simulation
exercise and ``real world'' needs for the medium energy beam transport (MEBT) of a
H$^-$-beamline.
Due to the space charge driven constraints the design gives only very little real estate 
for beam diagnostics.
However, neither the given space will be sufficient to accommodate all the
necessary instrumentation for beam intensity, orbit, phase, tr.\ and
long.\ emittances, as well as beam halo and tails measurements, nor
a single measurement location gives valuable information about the
beam parameters along the beamline, like a simulation program does.

\begin{figure}[h]
\centering
\includegraphics[width=80mm, clip=true]{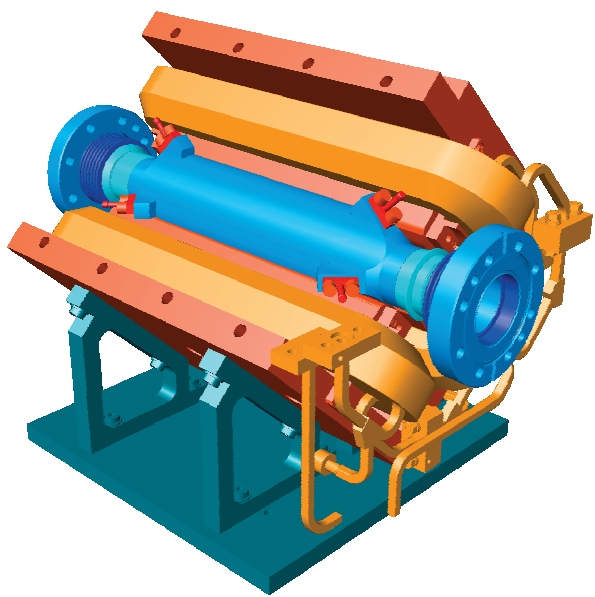}
\caption{BPM pickup inside a quadrupole saves space!} \label{BPMquad_figure}
\end{figure}

\emph{Space for beam diagnostics} was and always will be an issues, therefore
the design of beam pickups with minimum space requirements is crucial.
Fig.~\ref{BPMquad_figure} shows a stripline BPM pickup, embedded inside a
quadrupole magnet, so it does not eat any additional space. 

\section{Beam Instrumentation}
Beam diagnostic and instrumentation systems are used for:
\subsection{Beam Characterization}
Most beam measurements and applications fall in this category:
\subsubsection{Beam intensity, bunch charge, beam current}
This is the most fundamental property to be measured in a particle accelerator,
i.e.\ how much beam do we have in the machine? A \emph{direct-current current-transformer}
(DCCT) is able to measure the DC contents of the beam, while \emph{toroidal transformers}
can give information on the number of particles in a bunch.
A \emph{wall current transformer} provides a very high bandwidth (up to ~10~GHz), and
can be also used for long.\ bunch profile characterization (hadrons), as timing electrode, 
or for beam phase measurements.
\subsubsection{Beam position, orbit, phase, energy, 
betatron / synchrotron tune, chromaticity, etc.}
The \emph{beam position monitors} (BPM) are the most powerful -- and most expensive --
beam instrumentation system in a particle accelerator.
Many BPM detectors are located along the beam-line, typically four or more per betatron 
oscillation period.
The BPMs offer much more than just the measurement of the beam orbit, they are essential in
almost every beam measurement, e.g. beam phase and energy, injection optimization, tune
measurement, dynamic aperture and lattice function measurements, etc.
During machine commissioning, but also later they are the most effective diagnostics for
troubleshooting and error analysis.
A \emph{Schottky detector} is related to a BPM, but optimized to sense the finite number
of particles in a beam in the frequency domain.
The Schottly monitor can used to measure beam tunes and emittances in a non-invasive way.
\subsubsection{Particle distribution, sliced beam / bunch parameters}
There is a verity of methods, intercepting and non-intercepting, to measure the transverse
or longitudinal beam profile, which leads to the beam emittance,  e.g.\ wire-scanners 
or flying wires, secondary emission monitors (SEM) based on multi-wires or foils, 
screen monitors based on fluorescence or transition radiation, and different styles of
non-invasive beam profile diagnostics (IPM, ODR, EOS, DMC, Schottky, laser wire, 
e-beam scanner, etc.).
\subsubsection{Beam losses, halo and tails}
\emph{Beam loss monitors} (BLM) are detection elements outside the 
vacuum chamber, which
are sensitive to particle showers (scintillation, ionization).
Similar to BPMs they are distributed along the beam-line, located at ``strategic''
locations, e.g.\ at the quadrupoles.
Typically the BLMs are part of a complex
machine protection system, which protects vacuum and other components
from uncontrolled beam losses.
In superconducting accelerators the latency of the BPM detection system is critical, 
to respond in time and prevent a quench or other major impacts.
However, often BLMs are also used to fine tune the beam orbit or in other
empirical machine optimization procedures. 
Some loss monitor systems (fiber optics based) allow a qualitative radiation dosimetry,
other technologies provide the total integrated loss along a beam-line (long ion chambers).

For high intensity beams, the monitoring of the transverse beam halo, 
and longitudinal beam
tails is of great interest, as off-core particles tend to get lost along the beam-line.
The vibrating-wire monitor (a temperature measurement) or other physical or laser wire
methods provide these features.
\subsection{Feedback Systems}
Some beam monitors, typically BPM detectors, are used in automatic feedback
systems (no human interface), e.g. orbit feedback, beam tune stabilization, damping of
instabilities, etc.
The feedback acts on magnets (quadrupoles), ejection and correction elements
(kickers, beam damping electrodes),
or on parts of the RF system (voltage), etc.
The loop of slow (sec-range) feedback systems is usually closed through the data-acquisition
and control system, particular if a deterministic response time is not critical.
For fast feedback systems ($\mu$sec...msec range) the latency of each element is critical
to prevent an instable operation.
While the FB system is able to improve the stability in the pass-band frequency range,
it may add unwanted noise outside the designed specifications. 
\subsection{Beam Monitors}
\begin{figure}[h]
\centering
\includegraphics[width=80mm, clip=true]{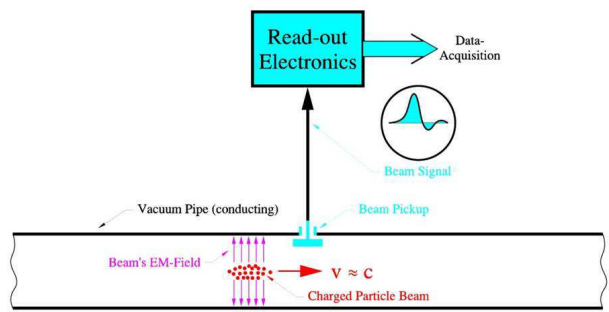}
\caption{Principle of a beam monitor.} \label{monitor_figure}
\end{figure}
Figure~\ref{monitor_figure} shows the simplistic principle of a
typical beam monitor, consisting out of two major elements:
\begin{description}
 \item[Beam detector] Typically the beam detector is part of 
 the vacuum system, and interacts with the beam in a non / minimum
 invasive or invasive way.
 As shown in the example (Fig.~\ref{monitor_figure}), the 
 electromagnetic field of  particle beam is sensed (here non-invasive) 
 by the detector and converted into an electric signal.
 Other minimum-invasive detection methods are based on the scattering with
 the residual gas or photons, synchrotron radiation detection, laser stripping
 of electrons (H$^-$-beams), or other ways of electromagnetic interaction.
 
 Invasive beam detectors use screen-, foil-, or wire-targets, and apply
 scintillation, secondary emission, or transition radiation principles.
 Cameras or charge detectors are used to convert to electrical signals.
 \item[Read-out system] The beam property of interest is embedded in the
 electrical signal provided by the detector.
 The read-out system extracts this information, e.g.\ bunch intensity,
 beam displacement, etc., and converts it a digital format which is accepted
 by the data-acquisition part of the accelerator control system.
 A read-out system can be as simple as a diode detector, or a very complex
 VME-crate with many modules.
 The trend is to convert the analog output signal of the detector
 as early as possible into a digital
 format, and make use of mathematical methods to extract the wanted beam
 information.
 Typically this digital signal processing is performed in a field programmable
 gate array (FPGA), which operates in close connection to an analog-to-digital
 converter (ADC). 
 
 Beside these core signal processing elements, 
 there are additional components, which are
 part of the read-out, acquisition \& control system of a beam monitor, e.g.\ 
 trigger, timing and control signals,
 power supplies, local control systems for switches, attenuators, motors and other
 motion control elements, cabling, internal interlocks and safety systems, etc.
\end{description}
\section{Requirements \& Challenges of Beam Diagnostics}
The requirements in beam quality and properties to be measured,
verified and controlled is challenging for any future HEP accelerator,
like a super-B factory, a high intensity hadron accelerator, a lepton
linear collider or a muon collider.
\begin{figure*}[t]
\centering
\includegraphics[width=135mm]{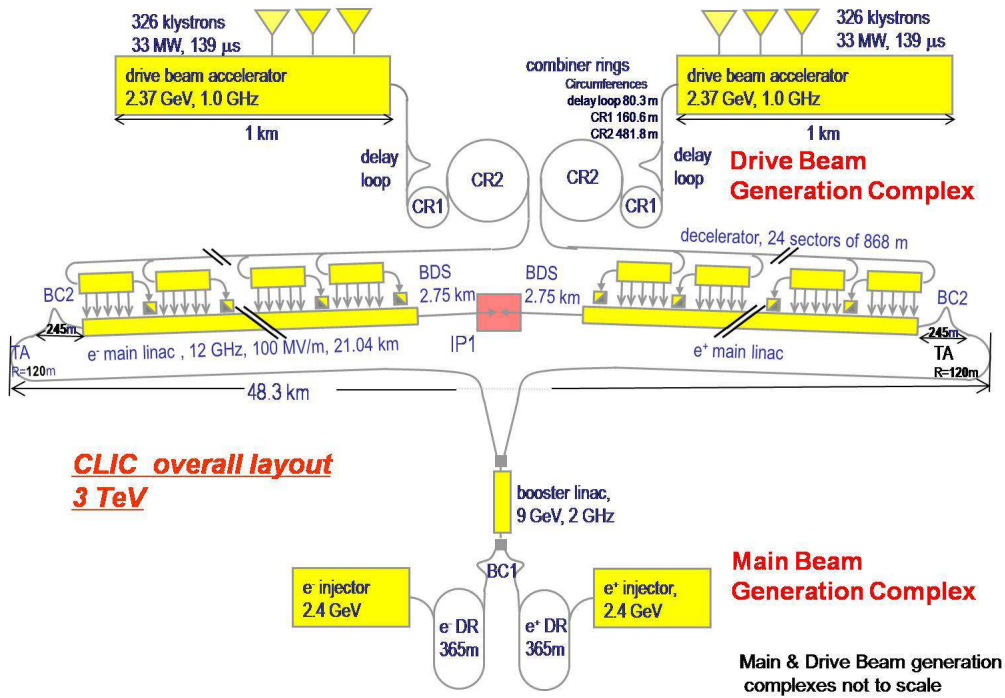}
\caption{The Compact Linear Collider (CLIC) HEP accelerator proposal.} \label{clic_figure}
\end{figure*}
Figure~\ref{clic_figure} shows the Compact Linear Collider (CLIC), proposed
by CERN as the next HEP energy frontier lepton machine.
The 3~TeV accelerator complex consists out of 96~km beam-lines, and
needs almost 200.000(!) beam monitor and diagnostic devices to observe and
control the beams.
Just the pure number of beam instrumentation elements is impressive (taking
about 10~\% of the total costs), the requirements of most systems is very
challenging, and at or beyond the current state-of-the-art.

The core requirements, e.g.\ resolution, reproducibility (and stability),
linearity, dynamic range, etc., for most of the beam diagnostic systems are 
established by beam dynamics simulations.
Other requirements and conditions are given by the available physical space,
environment (radiation, temperature, remote location), or laboratory standards
(data-acquisition formats, timing and trigger standards, rack and crate standards).
For complex and operational important instrumentation systems the reliability
plays a role.

Some key requirements for the beam diagnostics are rather different between
lepton and hadron accelerators:
\subsection{Lepton Accelerators -- \newline LCLS, XFEL, ILC, CLIC}
\begin{itemize}
 \item Observation and control of the longitudinal beam dynamics is the most
 challenging aspect for the beam instrumentation of these machines.
 The bunch length is in the 50\ldots500~fsec range, and requires very a high
 bandwidth (THz range) of the instruments to measure the longitudinal bunch profile.
 \item The BPM requirements are very challenging, i.e.\ a RMS resolution in
 the 50\ldots500~nm is needed, while the measurement (integration) time is
 in the 50\ldots500~nsec range.
 \item The large number of beam instruments (CLIC: ~200.000 beam monitors,
 thereof ~50.000 BPMs with sub-micrometer resolution) requires massive simplification
 (costs) and optimization for series production, testing, installation and system
 commissioning.
\end{itemize} 
\subsection{Hadron Accelerators -- \newline
 SNS, LHC, J-PARC, Project X, $\mu$-Collider}
\begin{itemize}
 \item For hadron machines the high beam power, and the related damage potential
 is in the foreground.
 Therefore the focus is on non-invasive beam monitors (laser wire, e-beam scanner,
 IPM).
 Many beam instruments are part of the machine protection system, which require
 a high reliability and/or a sufficient redundancy of these devices.
 \item The mitigation of beam losses is crucial.
 The characterization/minimization of beam halo and tails in the low
 and medium energy  
 beam transport (LEBT/MEBT) of these accelerators will prevent beam losses and activation
 in the high energy section of the machines.
 \item Special hadron beam diagnostics have to be established for the low-energy,
 non-relativistic parts of the accelerator.
 CW operation of hadron linacs will challenge the data-acquisition throughput and 
 time stamping requirements.
\end{itemize}
\section{Examples of Advanced Beam Instrumentation}
\subsection{High Resolution BPMs}
\begin{figure}[h]
\centering
\includegraphics[width=80mm, clip=true]{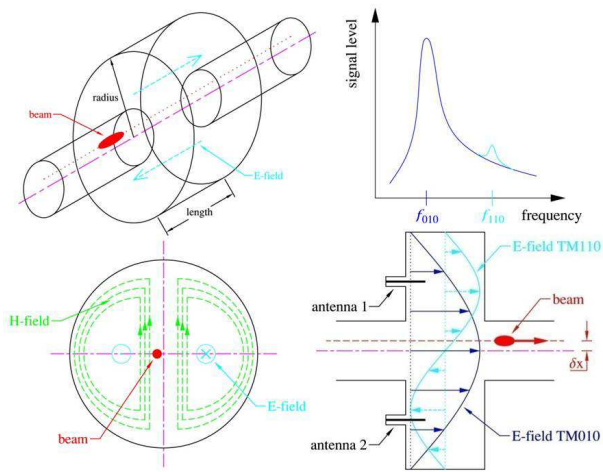}
\caption{High resolution resonant cavity BPM principle.} \label{cavityBPM_figure}
\end{figure}

To achieve a sufficient luminosity in the next generation
linear collider (e.g.\ ILC, CLIC), the vertical RMS beam size at
the IP has to be squeezed to just a few nm.
This requires a very accurate observation and control of the
beam orbit throughout the entire accelerator complex.
High resolution ($<$~1~$\mu$m) beam position monitors with
single-pass, single-bunch measurement capability are a key requirement.
Instead of sensing the mirror current distribution on the vacuum
chamber, induced by the charged particle beam (indicated in Fig.~\ref{monitor_figure}),
a beam excited dipole eigenmode (TM$_{110}$) of a passive resonant cavity gives an
intrinsic higher sensitivity to the beam displacement (Fig.~\ref{cavityBPM_figure}).

The resolution ``world-record'' is currently held by an Asian
team of beam instrumentation experts.
Based on a common-mode free C-Band ($\approx$6~GHz) rectangular cavity design
a resolution of 8.7~nm could be verified by beam measurements!
\subsection {Digital Signal Processing}
\begin{figure}[h]
\centering
\includegraphics[width=80mm, clip=true]{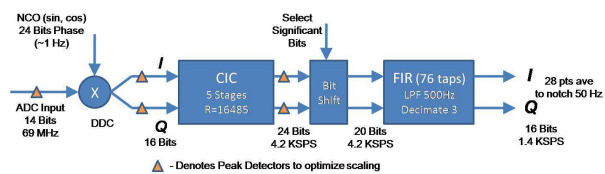}
\caption{Narrowband BPM signal processing.} \label{narrowband_figure}
\end{figure}

Beam orbit and damping systems gained the highest profit from
recent advances in the area of digital signal processing.
The generation and preservation of a low vertical emittance lepton beam
is a key element of any version of the next linear HEP collider,
and is mainly defined in the damping ring.
High resolution beam position monitors, based on simple ``button'' style electrodes,
allow to steer the beam along a ``golden'' orbit, with minimum non-linear
field effects.
\begin{figure}[h]
\centering
\includegraphics[width=80mm, clip=true]{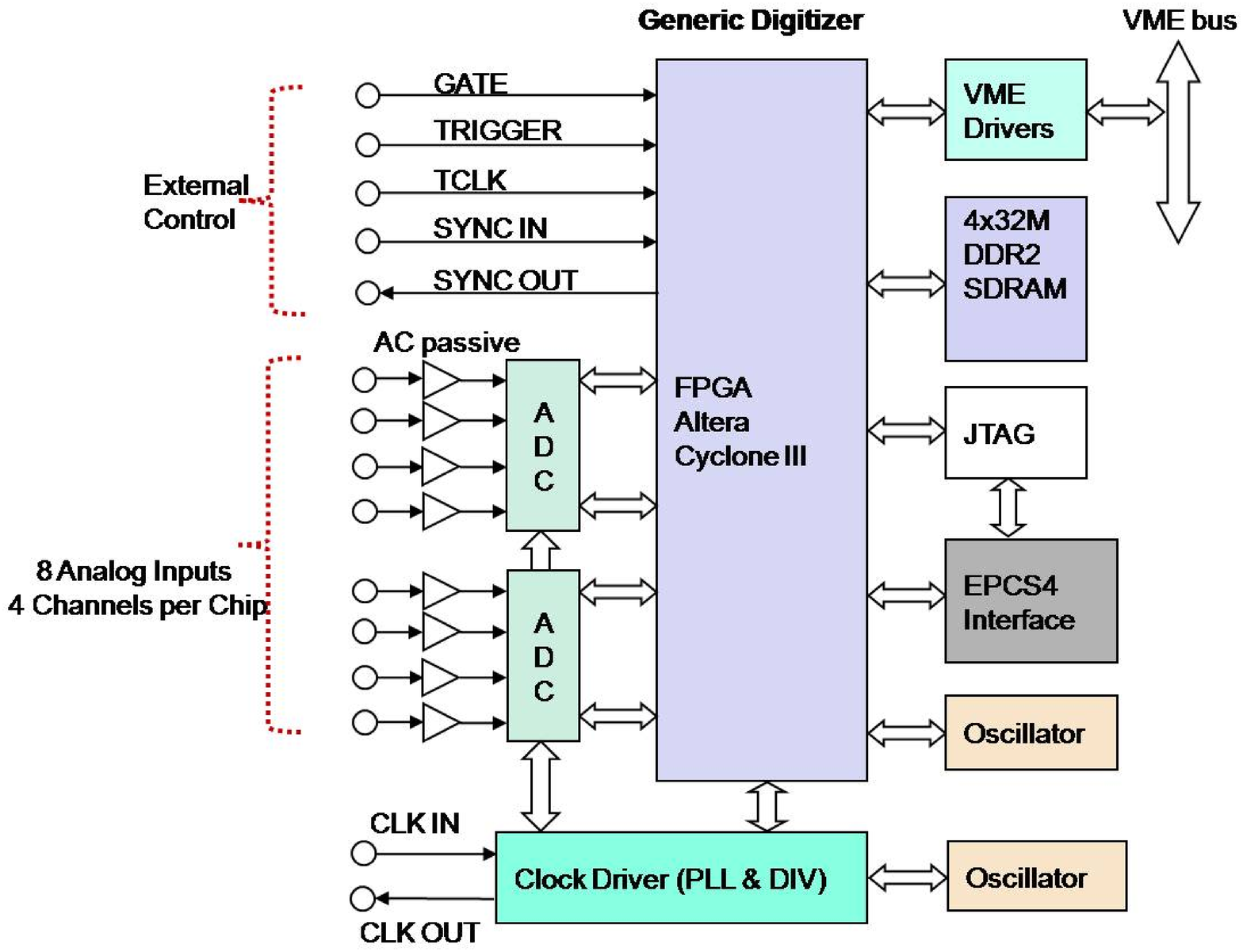}
\caption{Block diagram of a digitizer.} \label{digitizer_figure}
\end{figure}
The required signal processing technology has two elements
\begin{description}
 \item[Hardware] A versatile digitizer, based on fast analog-to-digital converters
 (ADC, typically 100-500 MSPS, 12-16 bit), a large field programmable gate array (FPGA)
 which holds all digital signal processing elements, and a low-jitter (10-500 fsec)
 timing and clock signal distribution system. 
 Figure~\ref{digitizer_figure} shows the block diagram of a VME digitizer.
 \item[Firmware] The FPGA firmware is used to downconvert, filter and decimate the
 data to extract the required beam position or orbit information.
 Typical digital building blocks are numerically controlled oscillators (NCO),
 downconverters (mixer), filters (FIR, IIR, CIC, etc.), delays, memory, mathematical
 functions, etc., even a complete microcomputer can be utilized.
 Figure~\ref{narrowband_figure} shows the narrowband signal processing section
 for a damping ring BPM system, which has a resolution potential of 100-200 nm.
\end{description}
\subsection{Optical Beam Diagnostics}
Most optical beam diagnostics are based on laser systems, and profit
from the availability of short (fsec range) laser pulses, in which 
the laser is mode-locked to the accelerator RF.
\subsubsection{Electro-Optical Sampling (EOS)}
\begin{figure}[h]
\centering
\includegraphics[width=80mm, clip=true]{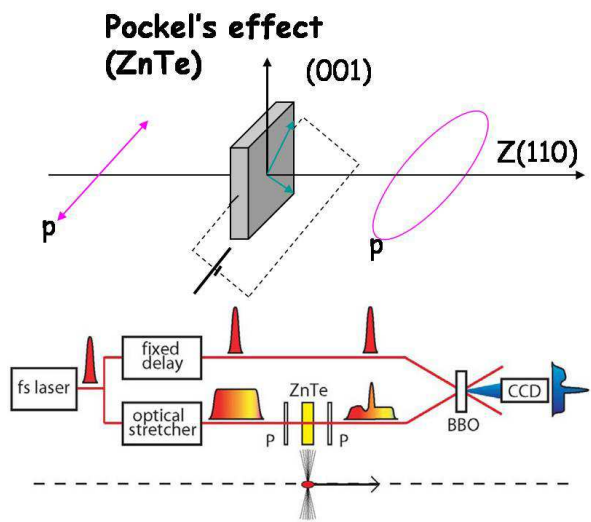}
\caption{\textit{Upper:} Pockel's cell effect.\newline
\textit{Lower:} Spectral-decoding electro-optical sampling.}
\label{eos_figure}
\end{figure}
The longitudinal bunch
profile of relativistic bunches can be measured in a non-invasive manner
with the \emph{electro-optical sampling} (EOS) technique.
Key element is a ZnTe crystal (\textit{Pockels} cell), which alters the polarization
of the transmitted laser light, while exposed to an electric field (upper part 
of Fig.~\ref{eos_figure}).
In an EOS setup the \textit{Pockels} cell is located inside the vacuum chamber, 
and sees the EM-field of the bunched beam, which alters the polarization of chirped
laser pulse -- imprinting its longitudinal profile.
The lower part of Figure~\ref{eos_figure} shows the details of this so-called
``spectral decoding'' EOS technique, which allows the single path profile measurement
of ultra-short bunches (resolution in the 50-500~fsec range).
\subsubsection{Laser wire}
\begin{figure}[h]
\centering
\includegraphics[width=80mm, clip=true]{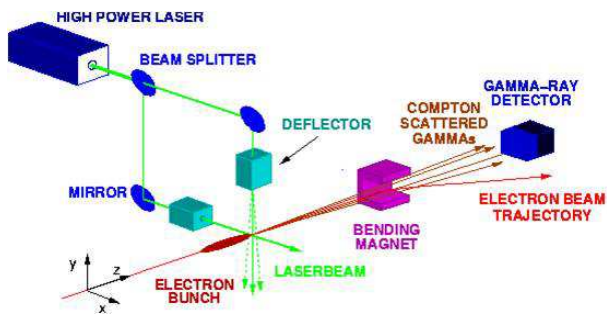}
\caption{Laser wire for lepton beams (compton scattering).}
\label{elaser_figure}
\end{figure}
The laser wire allows a non-invasive measurement of the transverse beam profile.
A thin physical wire scanning an electron or H$^-$ 
beam is replaced by a focused laser beam (Fig.~\ref{elaser_figure}), thus no wire heating or damage in case of high beam intensities, and
no fragile wire setup.
In case of an H$^-$ beam, the weakly bounded electrons (0.75 eV) are stripped: 
$$
H^-\,+\,\gamma\;\rightarrow\;H^0\,+\,e^-
$$
and with help of a dipole magnet separated from the H$^-$ beam for detection
(\textit{Faraday}-cup, scintillator-PMT, etc.).

To enhance the resolution of the laser wire, i.e.\ to analyze sub-micrometer
beam-sizes, the laser beam can be operated at a higher moment in an optical
cavity.
Experiments demonstrated a resolution of 4.3~$\mu$m, operating a laser beam
of 9.6~$\mu$m RMS size in the TEM$_{01}$ (dipole) mode.

\subsection{Other Advanced Beam Diagnostics}
Beside the few examples given, the a large variety of advanced
beam diagnostics is available to better understand the beam parameters, identify
deficiencies, locate errors and improve the beam quality.
\begin{description}
 \item[Transition or diffraction radiation] is used in the infrared, or optical
 domain to characterize lepton beam / bunch parameters, e.g. transverse profile,
 bunch length / profile, beam divergence / emittance, etc.
 The detection techniques, operating in the THz range (\emph{Golay}-cells,
 ``pyro''-detectors), include interferometric methods 
 (\emph{Michelson, Martin-Puplett}).
 \item[Electron Beam Scanner] is a non-invasive approach to measure the beam profile
 of a high intensity proton beam, i.e.\ scanning an e-beam through the p-beam
 and detect the deflection effect.
 A different non-invasive method is based on the ionization of the residual
 gas, the \textbf{ionization profile monitor} (IPM).
 \item[HOM, e-Cloud] We can utilize some beam detectors in
 ``other'' ways, e.g.\ we can used a set of two BPMs to measure the \emph{electron
 cloud} effect or instability under high beam intensities, 
 we can use the \emph{HOM-coupler signals} 
 of a RF accelerating structure to measure the beam displacement (or cavity alignment)
 to a few $\mu$m, or the beam phase to 0.1$^0$, and
 there are more examples of ``parasitic'' beam diagnostics.
\end{description}

\end{document}